\documentclass[aps,prb,a4paper,preprint,superscriptaddress,nobibnotes]{revtex4-1}
\usepackage{helvet}
\usepackage{graphicx}
\usepackage{xspace}
\usepackage{amsmath,amssymb}
\usepackage[usenames,dvipsnames]{xcolor}
\usepackage{multibib}
\newif\ifdrafttext
\drafttexttrue 
\ifdrafttext \usepackage[colorlinks,urlcolor=black,citecolor=black,linkcolor=black]{hyperref} \else \fi

\newcounter{myfigure}
\makeatletter
\@addtoreset{figure}{myfigure}
\def\@caption@fignum@sep{~$\vert$~}%
\def\fnum@figure{\textbf{\figurename~\thefigure}}
\makeatother

\newcommand{\ket}[1]{\ensuremath{|#1\rangle}}
\newcommand{\Bzdot}{\ensuremath{\dot{B}_z}}
\newcommand{\micro}{\ensuremath{\mu}}
\newcommand{\micron}{\ensuremath{\mu\mathrm{m}}\xspace}
\newcommand{\del}{\ensuremath{\nabla}}
\newcommand{\mb}[1]{\mathbf{#1}}
\renewcommand{\mb}[1]{#1}
\newcommand{\uv}[1]{\mb{\hat{#1}}}
\newcommand{\Afield}{\ensuremath{{A^*}}\xspace}
\newcommand{\Bfield}{\ensuremath{{B^*}}\xspace}
\newcommand{\Efield}{\ensuremath{{E^*}}\xspace}
\newcommand{\Atwiddle}{\ensuremath{{\tilde{A}}^*}\xspace}
\newcommand{\Vorticity}{\ensuremath\mathit{{\Omega}}}

\newcommand{\Bzf}{\ensuremath{B_{z,\mathrm{f}}}\xspace}

\newcommand{\Bqprime}{\ensuremath{b}_\mathrm{q}}

\newcommand{\Pad}{\ensuremath{\mathit{\Pi}_\mathrm{ad}}}
\newcommand{\mulocal}{\ensuremath{\mu}}

\newcommand{\abstracta}[1]{\textcolor{red}{#1}}
\newcommand{\abstractb}[1]{\textcolor{blue}{#1}}
\newcommand{\abstractc}[1]{\textcolor{ForestGreen}{#1}}
\newcommand{\abstractd}[1]{\textcolor{Fuchsia}{#1}}
\newcommand{\abstracte}[1]{\textcolor{Magenta}{#1}}
\newcommand{\abstractf}[1]{\textcolor{SkyBlue}{#1}}
\newcommand{\twiddle}[1]{\ensuremath{\tilde{#1}}}
\renewcommand{\boldsymbol}[1]{#1}
\newcommand\Phitilde{\stackrel{~\widetilde{}}{\smash{\mathit{\Phi}}\rule{0pt}{0.65ex}}}

    \renewcommand{\abstracta}[1]{#1}
    \renewcommand{\abstractb}[1]{#1}
    \renewcommand{\abstractc}[1]{#1}
    \renewcommand{\abstractd}[1]{#1}
    \renewcommand{\abstracte}[1]{#1}
    \renewcommand{\abstractf}[1]{#1}

\newcites{si}{References}
\newcommand*{\citensi}[1]{%
  \begingroup
    \romannumeral-`\x 
    \setcitestyle{numbers}%
    \citesi{#1}%
  \endgroup
}
\newcommand*{\citen}[1]{%
  \begingroup
    \romannumeral-`\x 
    \setcitestyle{numbers}%
    \cite{#1}%
  \endgroup
}

\begin{document}


\title{Observation of Dirac Monopoles in a Synthetic Magnetic Field}


\author{M. W. Ray}
\affiliation{Department of Physics, Amherst College, Amherst, Massachusetts 01002--5000, USA}

\author{E. Ruokokoski}
\affiliation{QCD Labs, COMP Centre of Excellence, Department of Applied Physics, Aalto University, P.O. Box 13500, FI--00076 Aalto, Finland}

\author{S. Kandel}
\altaffiliation[Present address: ]{City of Hope, 1500 East Duarte Road, Duarte, California 91010, USA.}
\affiliation{Department of Physics, Amherst College, Amherst, Massachusetts 01002--5000, USA}

\author{M. M\"ott\"onen}
\affiliation{QCD Labs, COMP Centre of Excellence, Department of Applied Physics, Aalto University, P.O. Box 13500, FI--00076 Aalto, Finland}
\affiliation{Low Temperature Laboratory (OVLL), Aalto University, P.O. Box 13500, FI-00076 Aalto, Finland}

\author{D. S. Hall}
\affiliation{Department of Physics, Amherst College, Amherst, Massachusetts 01002--5000, USA}

\date{20 September 2013; accepted 4 December 2013}

\preprint{\href{http://www.nature.com/nature/journal/v505/n7485/full/nature12954.html}{M. W. Ray et al., Nature {\bf 505}, 657 (2014)}; \href{http://dx.doi.org/10.1038/nature12954}{doi:10.1038/nature12954}}

\begin{abstract}
\abstracta{Magnetic monopoles --- particles that behave as isolated north or south magnetic poles --- have been the subject of speculation since the first detailed observations of magnetism several hundred years ago~\cite{Goldhaber1990}. Numerous theoretical investigations and hitherto unsuccessful experimental searches~\cite{Milton2006} have followed Dirac's 1931 development of a theory of monopoles consistent with both quantum mechanics and the gauge invariance of the electromagnetic field~\cite{Dirac1931}.} \abstractb{The existence of even a single Dirac magnetic monopole would have far-reaching physical consequences, most famously explaining the quantization of electric charge~\cite{Dirac1931,Vilenkin1994}.} \abstractc{Although analogues of magnetic monopoles have been found in exotic spin-ices~\cite{Castelnovo2008,Morris2009}
and other systems~\cite{Chuang1991,Fang2003,Milde2013}, there has been no direct experimental observation of Dirac monopoles within a medium described by a quantum field, such as superfluid helium-3 (refs~\citen{Blaha1976,Volovik1976,Salomaa1987,Volovik2003}).} \abstractd{Here we demonstrate the controlled creation~\cite{Pietila2009} of Dirac monopoles in the synthetic magnetic field produced by a spinor Bose-Einstein condensate.} \abstracte{Monopoles are identified, in both experiments and matching numerical simulations, at the termini of vortex lines within the condensate.  By directly imaging such a vortex line, the presence of a monopole may be discerned from the experimental data alone.} \abstractf{These real-space images provide conclusive and long-awaited experimental evidence of the existence of Dirac monopoles. Our result provides an unprecedented opportunity to observe and manipulate these quantum-mechanical entities in a controlled environment.}
\end{abstract}

\pacs{}

\maketitle


Maxwell's equations refer neither to magnetic monopoles nor to the magnetic currents that arise from their motion. Although a simple
symmetrisation
with respect to the electric and magnetic fields, respectively ${E}$ and ${B}$, leads to equations that involve these magnetic charges, it also seemingly prevents their description in terms of the familiar scalar and vector potentials, respectively $V$ and ${A}$, alone. Because the quantum-mechanical Hamiltonian is expressed in terms of potentials, rather than electromagnetic fields, this modification immediately leads to serious theoretical challenges. 

In a celebrated paper that combined arguments from quantum mechanics and classical electrodynamics~\cite{Dirac1931}, Dirac identified electromagnetic potentials consistent with the existence of magnetic monopoles. His derivation relies upon the observation that in quantum mechanics the potentials $V$ and $\mb{A}$ influence charged particle dynamics either through the Hamiltonian or, equivalently, through modifications of the complex phase of the particle wavefunction. Armed with these equivalent perspectives, Dirac then considered the phase properties of a wavefunction pierced by a semi-infinite nodal line with nonzero phase winding. He discovered that the corresponding electromagnetic potentials yield the magnetic field of a monopole located at the endpoint of the nodal line. The vector potential $\mb{A}$ in this case also exhibits a nonphysical line singularity, or `Dirac string', that terminates at the monopole. 

\begin{figure}[p!,floatfix]
\ifdrafttext \includegraphics[width=\linewidth]{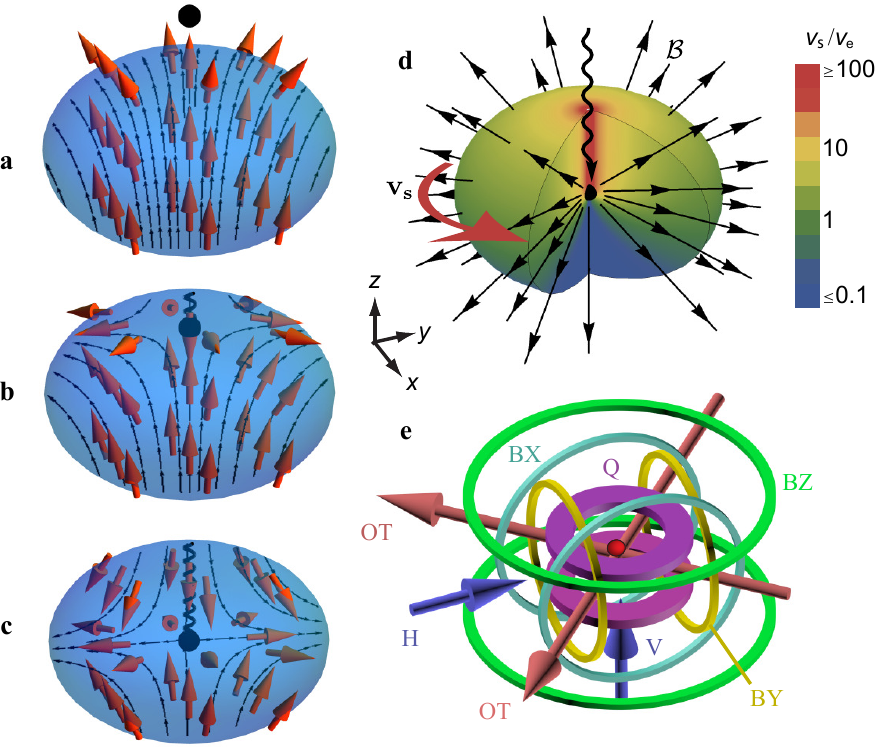} \else \fi
\caption{\label{fig:monopole-overview}\textbf{Schematic representations of the monopole creation process and experimental apparatus.} \textbf{a-c}, Theoretical spin orientation (red arrows) within the condensate when the magnetic field zero (black dot) is above (\textbf{a}), entering (\textbf{b}) and in the middle of (\textbf{c}) the condensate.  The helix represents the singularity in the vorticity. \textbf{d}, Azimuthal superfluid velocity $v_s$ (colour scale and red arrow), scaled by equatorial velocity $v_e$. Black arrows depict the synthetic magnetic field, $\Bfield$. \textbf{e}, Experimental setup showing magnetic quadrupole (Q) and bias field (BX, BY and BZ) coils.  Red arrows (OT) show beam paths of the optical dipole trap, and blue arrows indicate horizontal (H) and vertical (V) imaging axes. Gravity points in the ${-}z$ direction.}
\end{figure}

We experimentally create Dirac monopoles in the synthetic electromagnetic field that arises in the context of a ferromagnetic spin-1 $^{87}$Rb Bose-Einstein condensate (BEC) in a tailored excited state~\cite{Pietila2009}. The BEC is described by a quantum-mechanical order parameter that satisfies a nonlinear Schr\"odinger equation, and the synthetic gauge potentials describing a north magnetic pole (Fig.~\ref{fig:monopole-overview}) are generated by the spin texture. This experiment builds on studies of synthetic electric and magnetic fields \Efield and \Bfield in atomic BECs, which is an emerging topic of intense interest in the simulation of condensed-matter systems with ultracold atoms~\cite{Lin2009,Dalibard2011}. Unlike monopole experiments in spin ices~\cite{Castelnovo2008,Morris2009}, liquid crystals~\cite{Chuang1991}, skyrmion lattices~\cite{Milde2013}, and metallic ferromagnets~\cite{Fang2003}, our experiments demonstrate the essential quantum features of the monopole envisioned by Dirac~\cite{Dirac1931}. 

\begin{figure}[p!,floatfix]
\ifdrafttext \includegraphics[width=3.5in]{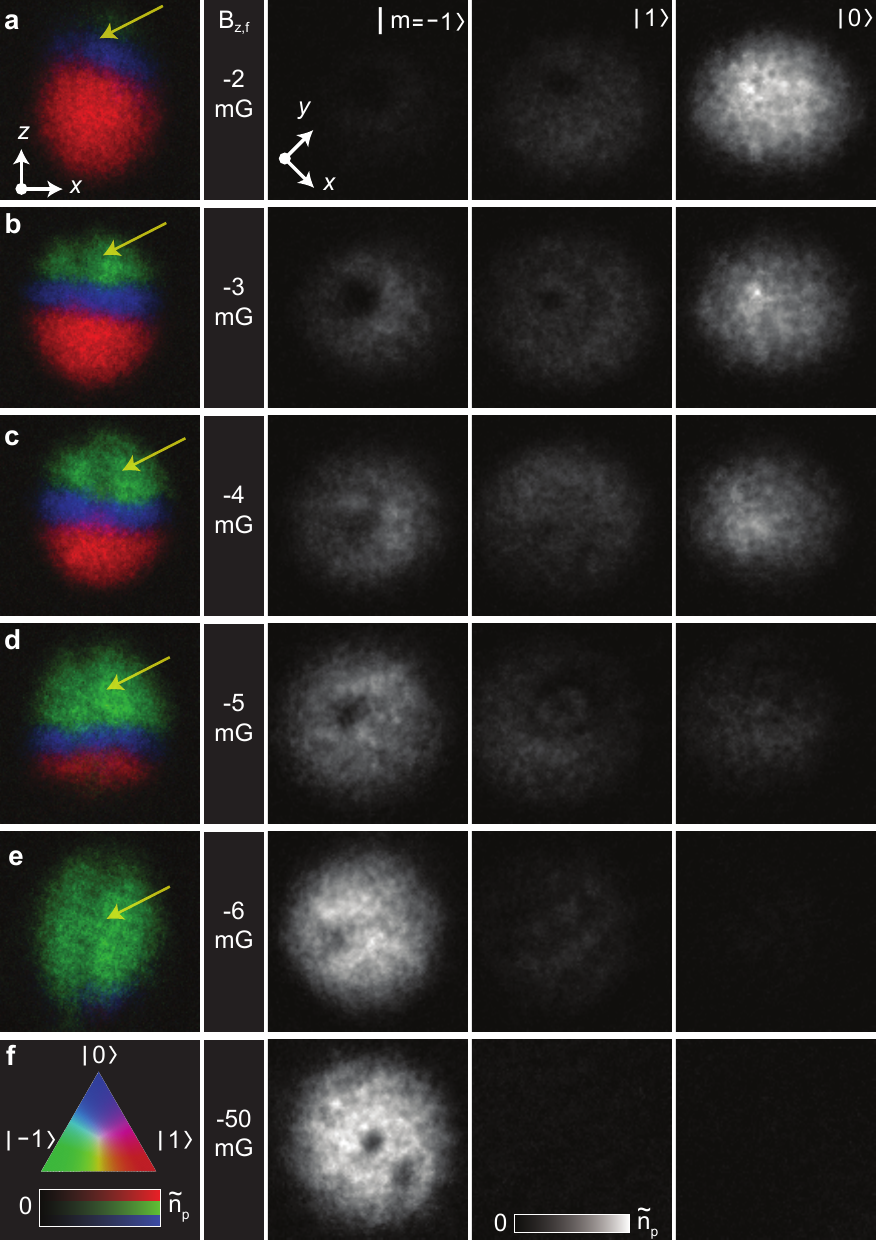} \else \fi
\caption{\label{fig:vortex_monopole_depth}\textbf{Experimental creation of Dirac monopoles.} Images of the condensate showing the integrated particle densities in different spin components as \Bzf\ is decreased. Each row contains images of an individual condensate. The leftmost column shows colour composite images of the column densities taken along the horizontal axis for the three spin states $\{\ket{1},\ket{0},\ket{{-}1}\}$; the colour map is given in \textbf{f}. Yellow arrows indicate the location of the nodal lines. The right three columns show images taken along the vertical axis. The scale is $285~\mu\mathrm{m} \times 285~\mu\mathrm{m}$ (horizontal) and $220~\mu\mathrm{m} \times 220~\mu\mathrm{m}$ (vertical), and the peak column density is $\tilde{n}_p = 1.0 \times 10^{9}$~cm${}^{-2}$.}
\end{figure}

\begin{figure}[p!,floatfix]
\ifdrafttext \includegraphics[width=3.5in]{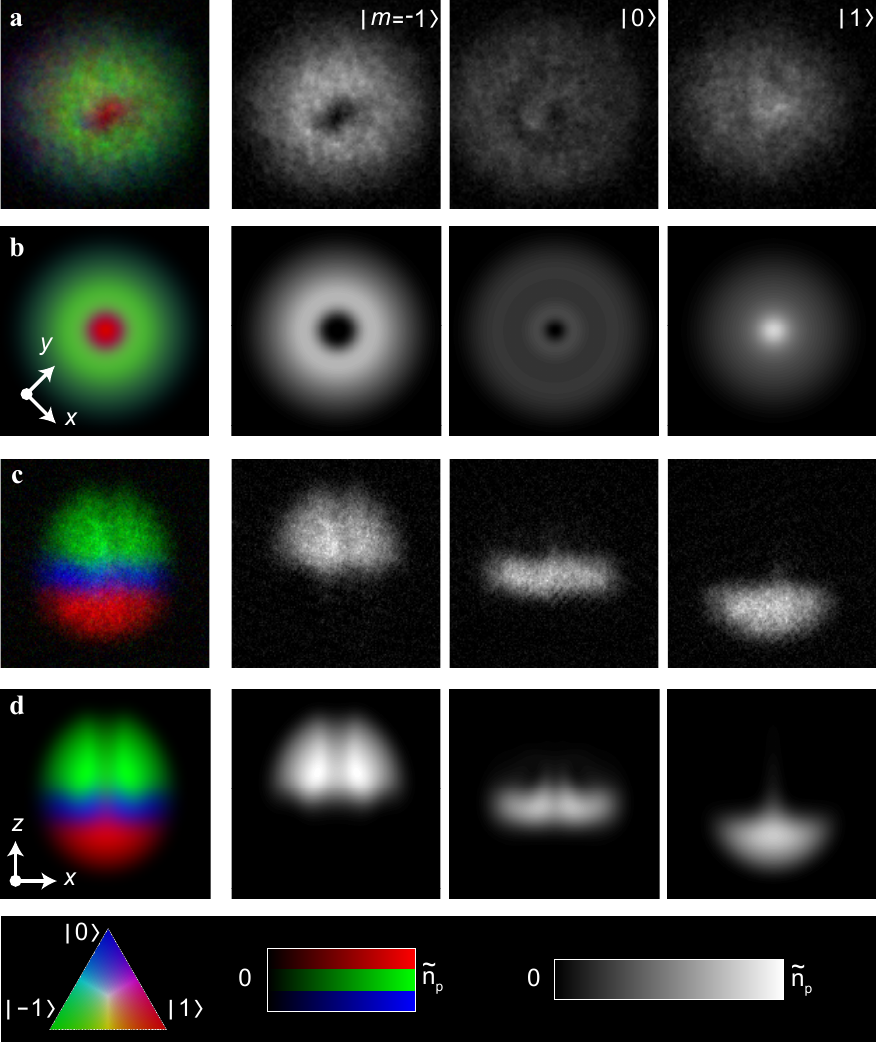} \else \fi
\caption{\label{fig:vortex_monopole-center}\textbf{Comparison between experiment and simulation.} Experimental (\textbf{a,c}) and simulated (\textbf{b,d}) condensate particle densities with the monopole near the centre of the condensate. Comparisons along the vertical axis are shown in rows \textbf{a} and \textbf{b}, while those along the horizontal axis are shown in rows \textbf{c} and \textbf{d}.  The hole observed in the \ket{{-}1} component (row \textbf{a}) is discernible as a line of diminished density in row \textbf{c}. The field of view is $220~\mu\mathrm{m} \times 220~\mu\mathrm{m}$ in \textbf{a} and \textbf{b} and $285~\mu\mathrm{m} \times 285~\mu\mathrm{m}$ in  \textbf{c} and \textbf{d}. The colour composite images and peak density $n_p$ are as in Fig.~\ref{fig:vortex_monopole_depth}.}
\end{figure}

\begin{figure}[p!]
\ifdrafttext \includegraphics[width=\linewidth]{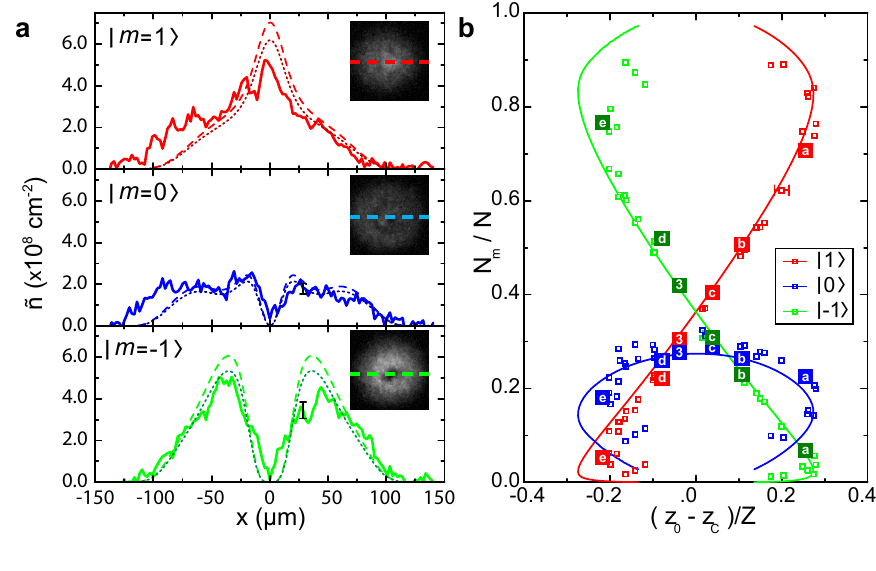} \else \fi
\caption{\label{fig:quantitative}\textbf{Quantitative comparison between  experiment and simulation.}
\textbf{a}, Experimental (solid lines) and simulated (dashed and dotted lines) column densities $\tilde{n}$ of the condensate from the vertical images in  Fig.~\ref{fig:vortex_monopole-center}, with cross-sections taken as shown in the insets. Dotted lines show the approximate effect of three-body losses (see text). The origin $x=0$ coincides with the hole in state $\ket{0}$. \textbf{b}, Fractions in each spin state for different  positions of the centre of mass of the \ket{0} state ($z_0$) relative to that of the condensate ($z_c$) in units of the axial Thomas-Fermi radius ($Z$). Solid lines are simulated values and points marked with letters and numbers correspond to panels \textbf{a-e} of Figs~\ref{fig:vortex_monopole_depth} and~\ref{fig:vortex_monopole-center}, respectively. Typical error bars that reflect uncertainties in the calibration of the imaging system are shown for several points.}
\end{figure}

Physically, the vector potential, $\Afield$, and synthetic magnetic field, $\Bfield = \hbar \del \times \Afield$, are related to the superfluid velocity, ${v_s}$, and vorticity, $\Vorticity = \del \times \mb{v_s}$, 
respectively. (Here $\hbar$ denotes Planck's constant divided by $2\pi$.) Our primary evidence for the existence of the monopole comes from images of the condensate density taken after the creation of these fields (Figs~\ref{fig:vortex_monopole_depth} and~\ref{fig:vortex_monopole-center}), which reveal a nodal vortex line with $4\pi$ phase winding terminating within the 
condensate. The images also display a three-dimensional spin structure that agrees well with the results of numerical simulations (Fig.~\ref{fig:quantitative}). We analyse these findings and discuss their implications below. 

The spinor order parameter corresponding to the Dirac monopole~\cite{Savage2003,Pietila2009} is generated by an adiabatic spin rotation in response to a time-varying magnetic field, ${B}({r},t)$. Similar spin rotations have been used to create multiply-quantized vortices~\cite{Leanhardt2002} and skyrmion spin textures~\cite{Choi2012a}. The order parameter $\mathit{\Psi}({r},t) = \psi({r},t) \zeta({r},t)$ is the product of a scalar order parameter, $\psi$ and a spinor, $\zeta = (\zeta_{+1},\zeta_0,\zeta_{-1})^\mathrm{T}\,\widehat{=}\,\ket{\zeta}$, where $\zeta_m=\langle m|\zeta\rangle$ represents the $m$th spinor component along $z$. The condensate is initially spin-polarised along the $z$~axis, that is, $\zeta = (1,0,0)^\mathrm{T}$. Following the method introduced in ref.~\citen{Pietila2009}, a magnetic field $\mb{B}(\mb{r},t) = \Bqprime(x\uv{x}+y\uv{y}-2z\uv{z})+B_z(t)\uv{z}$ is applied, where $\Bqprime>0$ is the strength of a quadrupole field gradient and $B_z(t)$ is a uniform bias field. The magnetic field zero is initially located on the $z$~axis at $z=B_z(0)/(2\Bqprime) \gg Z$, where $Z$ is the axial Thomas-Fermi radius of the condensate. The spin rotation occurs as $B_z$ is reduced, drawing the magnetic field zero into the region occupied by the superfluid.

Ideally, the condensate spin adiabatically follows the local direction of the field (Fig.~\ref{fig:monopole-overview}{a}--{c}). Our numerical analysis indicates, and both simulations and experiment confirm, that the fraction of atoms undergoing nonadiabatic spin-flip transitions is of order $1\%$ for our experimental parameters. The spin texture in the adiabatic case is conveniently expressed in a scaled and shifted coordinate system with $x'=x$, $y'=y$, $z'=2z-B_z/\Bqprime$, corresponding derivatives $\nabla'$, and spherical coordinates $(r',\theta',\varphi')$. This transformation scales the $z$~axis by a factor of two and shifts the origin of coordinates to coincide with the magnetic field zero. The applied magnetic field is then ${B} = \Bqprime(x'\uv{x}' + y'\uv{y}' - z'\uv{z}')$. As $B_z$ is reduced, each spin rotates by an angle $\pi-\theta'$ about an axis defined by the unit vector $\uv{n}(r',\theta',\varphi')= -\uv{x}'\sin\varphi' + \uv{y}'\cos\varphi' \label{eq:n}$. This spatially-dependent rotation leads to a superfluid velocity
\begin{align}
{v_s} = \frac{\hbar}{Mr'}\frac{1+\cos\theta'}{\sin\theta'}{{\hat{\varphi'}}} \label{eq:velocity}
\end{align}
and vorticity
\begin{align}\label{eq:vorticity}
\Vorticity = 
-\frac{\hbar}{M r'^2}\mb{\hat{r'}} + \frac{4 \pi \hbar}{M} \delta(x') \delta(y') \mathit{\Theta}(z') \mb{\hat{r'}}
\end{align}
where $M$ is the atomic mass, $\delta$ is the Dirac delta function and $\mathit{\Theta}$ is the Heaviside step function. The vorticity is that of a monopole attached to a semi-infinite vortex line singularity, of phase winding $4\pi$, extending along the positive $z'$~axis.

The synthetic vector potential arising from the spin rotation can be written as $\Afield = - M {v_s}/\hbar$, with the line singularity in $\Afield$ coincident with the nodal line in $\mathit{\Psi}$. However, this singularity is nonphysical, as it depends on the choice of gauge and can even be made to vanish~\cite{Wu1975} (Supplementary Information). The synthetic magnetic field of the monopole is therefore simply
\begin{align}\label{eq:monopolefield}
\Bfield = \frac{\hbar}{r'^2}\,\uv{r}'
\end{align}
The fields ${v_s}$ and $\Bfield$ are depicted in Fig.~\ref{fig:monopole-overview}{d}.

The experimental setup~\cite{Kaufman2009} is shown schematically in Fig.~\ref{fig:monopole-overview}\textbf{e}. The optically-trapped $^{87}$Rb BEC consists of $N = 1.8(2)\times 10^5$ atoms in the $\ket{F{=}1,m{=}1}\equiv\ket{1}$ spin state, where the uncertainty reflects shot-to-shot variations and the calibration of the detection system. The calculated radial and axial Thomas-Fermi radii are $R=6.5~\mu$m and $Z = 4.6~\mu$m, respectively, and the corresponding optical trap frequencies are respectively $\omega_4 \approx 2\pi \times 160$~Hz and $\omega_z \approx 2\pi \times 220$~Hz. Four sets of coils are used to produce $b_q$, $B_z$ and the transverse magnetic field components $B_x$ and $B_y$, which are used to guide the applied magnetic field zero into the condensate. At the beginning of the monopole creation process, the bias field is $B_z = 10$~mG. The quadrupole field gradient is then linearly ramped from zero to $\Bqprime = 3.7$~G/cm, placing the magnetic field zero approximately $30~\micron$ above the condensate. The field zero is then brought down into the condensate by decreasing $B_z$ linearly to $\Bzf$ at the rate $\Bzdot = -0.25$~G/s. We call this the creation ramp.

\begin{figure}[p!]
\ifdrafttext\includegraphics[width=\linewidth]{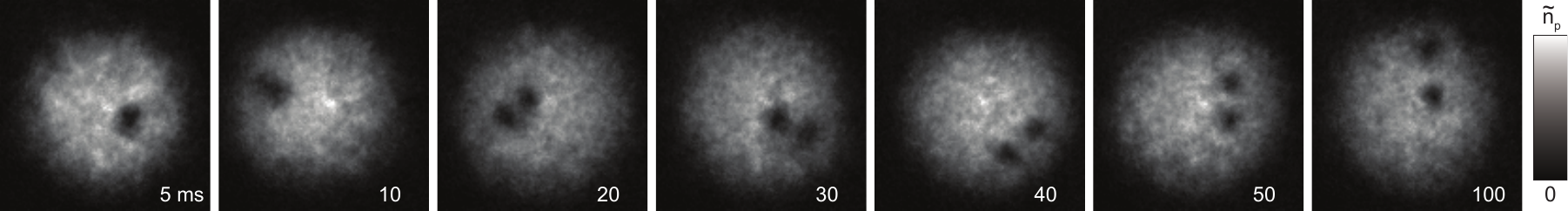} \else \fi %
\caption{\label{fig:vortexsplit}\textbf{Decay of the doubly-quantized vortex.} Images of the condensate time-evolution after moving the magnetic field zero completely through the condensate. The evolution time is shown at the bottom right of each panel.  The maximum pixel intensity corresponds to a peak column density $\tilde{n}_p = 1.0 \times 10^{9}~\mathrm{cm}^{-2}$, and the field of view is $246~\mu\mathrm{m} \times 246~\mu\mathrm{m}$.  Each image represents a separate condensate, and $\Bzdot = 3$~G~s$^{-1}$.  After roughly 10~ms the vortex splits in two, demonstrating the initial 4$\pi$ phase winding of the nodal line.}
\end{figure}

The atomic density of each spinor component $\ket{m}$ is imaged as established by the local spin rotation during the creation ramp (Methods). As the field zero passes through the condensate (Fig.~\ref{fig:vortex_monopole_depth}{a}--{f}), the distribution of particles in the three spin states changes in a manner indicative of the expected spin rotation shown in Fig.~\ref{fig:monopole-overview}. The nodal line appears in the images taken along the vertical axis as holes in the \ket{{-}1} and $\ket{0}$ components, and in the side images as regions of reduced density extending vertically from the top of the condensate towards, but not through, the $\ket{1}$ component. This nodal line extends more deeply into the condensate as \Bzf is reduced. Ultimately it splits into two vortex lines (Fig.~\ref{fig:vortex_monopole_depth}{f}; see also Extended Data Fig.~\ref{fig:vortexsplit}) --- the characteristic signature of the decay of a doubly-quantized vortex~\cite{Shin2004} --- illustrating its $4\pi$ phase winding.

We compare the experimental images of the vertically (Fig.~\ref{fig:vortex_monopole-center}{a}) and horizontally (Fig.~\ref{fig:vortex_monopole-center}{c}) imaged density profiles to those given by numerical simulations (Fig.~\ref{fig:vortex_monopole-center}{b},{d}) in which the monopole is near the centre of the condensate. The simulation data are obtained by solving the full three-dimensional dynamics of the spinor order parameter (Methods). The locations of the doubly-quantized and singly-quantized vortices in spinor components $\ket{{-}1}$ and $\ket{0}$ are clearly visible in the experimentally acquired density profiles, as are other structures discernible in the images obtained from the numerical simulations. The observed vertical spatial separation of the spinor components,
(Fig.~\ref{fig:vortex_monopole-center}{c}) confirms that the vortex line terminates within the bulk of the condensate.

The quantitative agreement between experiment and simulation is apparent in
Fig.~\ref{fig:quantitative}, which shows cross-sections of the density profiles taken through the centre of the condensate. The differences observed in the peak densities (Fig.~\ref{fig:quantitative}{a}) of experimental (solid lines) and simulated (dashed lines) data are due to effects not taken into account in the simulation, such as three body losses that were observed to be $\sim 10 \%$ in the experiment.  To show their effect we have scaled the simulated data accordingly (dotted lines).  Noting the absence of free parameters, the experimental data are in very good agreement with the numerical simulation.

We also show the fraction of the condensate in each spinor component for different vertical monopole locations within the condensate (Fig.~\ref{fig:quantitative}{b}), including data from images in which the nodal line of the order parameter does not necessarily
coincide with the $z$~axis. The physical observable is the position of the centre of mass of the $\ket{0}$ component, $z_0$, relative to the centre of mass of the whole condensate, $z_c$. Again, we find the experiments and simulations are in very good quantitative agreement without any free parameters.

An alternative description of the origins of the velocity and vorticity profiles (equations~\eqref{eq:velocity} and~\eqref{eq:vorticity}) can be presented in terms of the motion of the monopole (Supplementary Information). As the monopole approaches the condensate, it is a source not only of the synthetic magnetic field \Bfield (equation~\eqref{eq:monopolefield}) but also of an azimuthal synthetic electric field, \Efield, described by Faraday's law, $\del' \times \Efield = - \partial \Bfield/\partial t$. Each mass element of the superfluid is given a corresponding azimuthal acceleration by \Efield. The monopole motion thereby 
induces the appropriate superfluid velocity and vorticity profiles within the condensate, in a manner similar to the induction of electric current in a superconducting loop by the motion of a (natural) magnetic monopole~\cite{Cabrera1982}. In our case, the condensate itself is the monopole detector, analogous to the superconducting loop. Being three-dimensional, however, it is sensitive to the entire $4\pi$ solid angle surrounding the monopole.

The creation and manipulation of a Dirac monopole in a controlled environment opens up a wide range of experimental and theoretical investigations. The time evolution and decay \cite{Pietila2009} of the monopole are of particular interest because it is not created in the ground state~\cite{Ruokokoski2011}. Interactions between the monopole and other topological excitations, such as vortices, present another fundamental research avenue with a variety of unexplored phenomena. There exists also the possibility of identifying and studying condensate spin textures that correspond to other exotic synthetic electromagnetic fields, such as that of the non-Abelian monopole~\cite{Pietila2009a}. Finally, the experimental methods developed in this work can also be directly used in the realization of a vortex pump~\cite{Mottonen2007}, which paves the way for the study of peculiar many-body quantum states, such as those related to the quantum Hall effect~\cite{Roncaglia2011}.

Note added. The effects of the Lorentz force arising from an inhomogeneous synthetic magnetic field have recently been observed in condensate dynamics~\cite{Choi2013}.

\begin{center}
\textbf{METHODS SUMMARY}
\end{center}

\textbf{Imaging.} After the creation ramp, we non-adiabatically change $B_z$ from $B_{z,\textrm{f}}$ to a large value (typically several hundred milligauss) in order to project the condensate spin components $\{\ket{m}\}$ into the approximate eigenstates of the Zeeman Hamiltonian while preserving the monopole spin texture. We call this the `projection ramp'. The condensate is then released from the trap and allowed to expand for $22.9$~ms.  The three spin states are separated along the $x$~axis during the expansion by a $3.5$~ms pulse of the magnetic field gradient with the magnetic bias field pointing in the $x$ direction. We take images simultaneously along the horizontal and vertical axes.

\begin{figure}[p!]
\ifdrafttext\includegraphics[width=\linewidth]{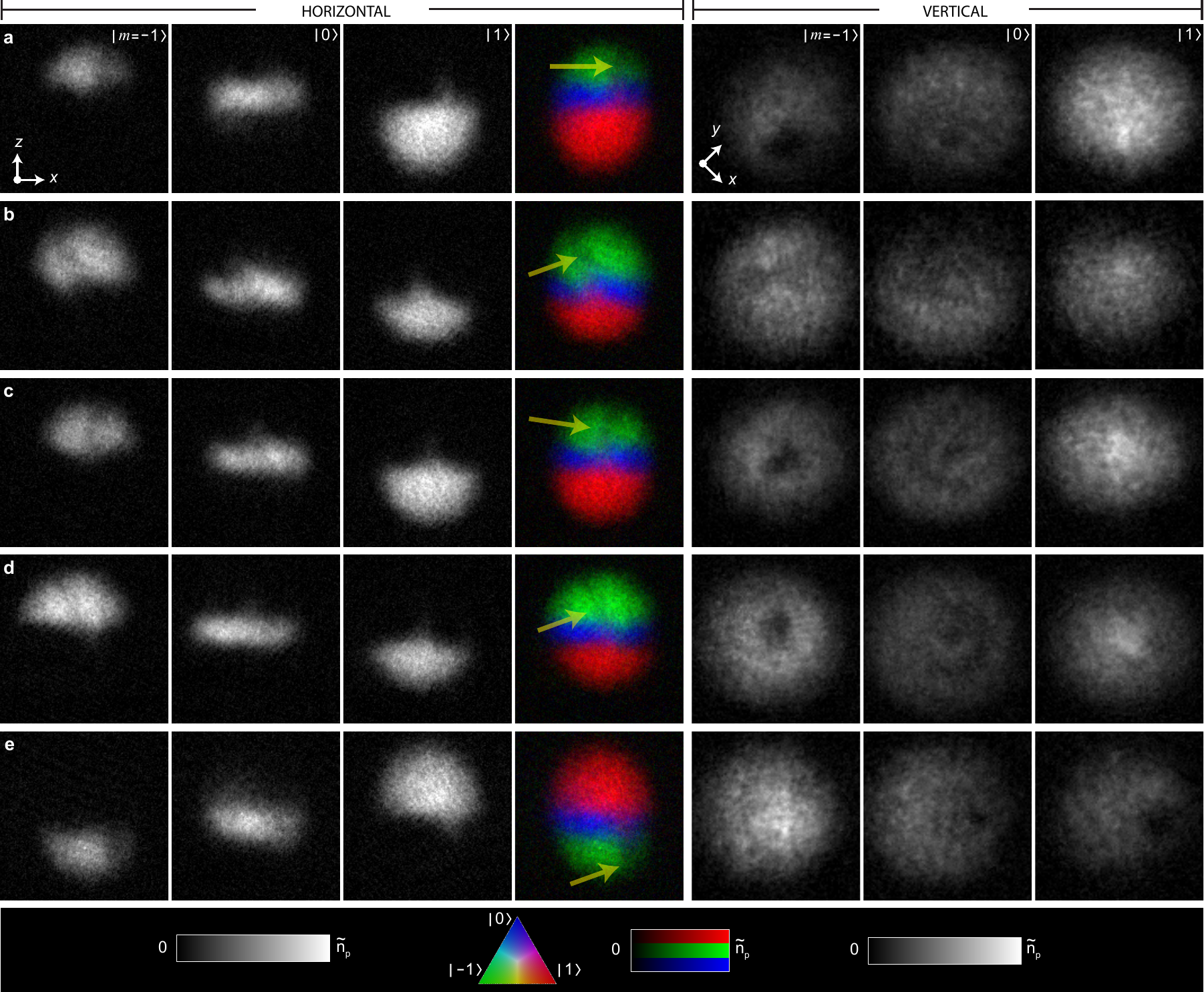} \else \fi %
\caption{\label{fig:extendeddata}\textbf{Additional representative images of Dirac monopoles.} Each row contains images of the same condensate.  The maximum pixel intensity corresponds to
$\tilde{n}_p = 8.2 \times 10^{8}~\mathrm{cm}^{-2}$, and the field of view is $220~\mu\mathrm{m}  \times 220~\mu\mathrm{m}$ in the vertical images, and $285~\mu\mathrm{m} \times 285~\mu\mathrm{m}$ in the horizontal images.  The arrow points to the density depletion that is identified as the nodal line.  In \textbf{a}--\textbf{c} we use the same protocol outlined in the paper: an off-centre monopole (\textbf{a}); an angled nodal line that is visible in the side image, but not in the vertically directed image (\textbf{b}); and a nodal line that appears to be splitting into two vortices in the \ket{m{=}{-1}} component (\textbf{c}). \textbf{d}, An example of a monopole spin structure in which the creation ramp is as described in the text but the projection ramp is reversed (that is, $B_z$ is rapidly increased until $|B_z/\Bzf| \gg 1$). \textbf{e}, Monopole spin structure created by moving the field zero into the condensate from below with $\Bzdot > 0$. The projection ramp is performed as described in \textbf{d}.}
\end{figure}

\textbf{Data.} The images shown in Figs~\ref{fig:vortex_monopole_depth} and~\ref{fig:vortex_monopole-center} are selected from among several dozen similar images taken under identical conditions, and hundreds of similar images taken under similar conditions (see also Extended Data Fig.~\ref{fig:extendeddata} for representative examples). Not every experimental run yields an image of a monopole, because drifts in the magnetic field and location of the optical trap cause the magnetic field zero to pass outside the BEC. Under optimal conditions, 5--10 consecutive images may be taken before drifts require adjustment of the bias fields.

\begin{figure}[p!]
\ifdrafttext\includegraphics[width=3.5in]{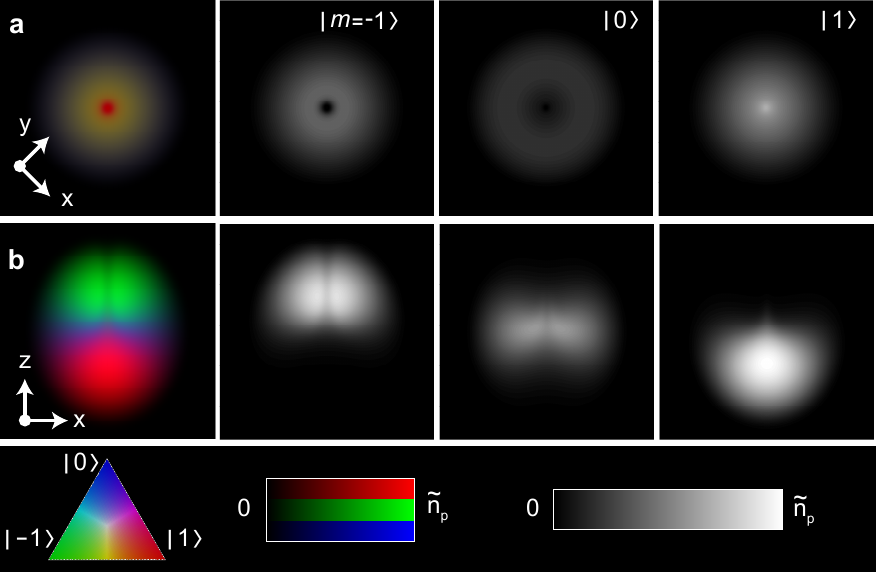} \else \fi %
\caption{\label{fig:supplement-simulation}\textbf{Numerical simulation of integrated particle densities before expansion.} Vertically (\textbf{a}) and horizontally (\textbf{b}) integrated particle densities of a condensate just before the projection ramp, with $\Bzf$ chosen such that the monopole is in the centre of the condensate. The fields of view are $17.2~\mu\mathrm{m} \times 17.2~\mu\mathrm{m}$ (\textbf{a}) and $17.2~\mu\mathrm{m} \times 11.4~\mu\mathrm{m}$ (\textbf{b}); in \textbf{b}, it is reduced in the $z$ direction for a more convenient comparison to the simulations shown in Fig.~\ref{fig:vortex_monopole-center}. The maximum pixel intensity corresponds to $\tilde{n}_p = 2.98 \times 10^{11}$ cm$^{-2}$.} 
\end{figure}

\textbf{Simulation.} We solve the full three-dimensional Gross--Pitaevskii equation with simulation parameters chosen to match those of the experiment, excepting the effects of three-body losses and the magnetic forces arising from the gradient during the spin component separation just before imaging. To show the effects of the expansion, we present integrated particle densities of the condensate from the numerical simulation immediately after the creation ramp, and while the magnetic field zero is still in the condensate, in Extended Data Fig.~\ref{fig:supplement-simulation}. The volume considered varies from $20\times 20\times 20 \, a_r^3$ to $320 \times 320\times 320 \, a_r^3$, where $a_r=\sqrt{\hbar/(M\omega_r)}\approx 0.9~\micron$ is the radial harmonic oscillator length. The size of the computational grid changes from $180\times 180\times 180$ to $1,024\times 1,024\times 1,024$ points.

\bibliographystyle{naturemag}
\bibliography{monopole}


Supplementary Information is available online.

\textbf{Acknowledgements} We acknowledge funding by the National Science Foundation (grants PHY--0855475 and PHY--1205822), by the Academy of Finland through its Centres of Excellence Program (grant no.\ 251748) and grants (nos\ 135794, 272806, and 141015), and ny Finnish Doctoral Programme in Computational Sciences. CSC - IT Center for Science Ltd.\ is acknowledged for computational resources (Project No.\ ay2090). We thank G.\ Volovik, M.\ Krusius, R.~H.\ Romer, M. Nakahara, and J.~R. Friedman for their comments on the manuscript. We also thank Heikka Valja for his artistic input. M.W.R.\ and D.S.H.\ acknowledge discussions with R.~P.\ Anderson, K. Jagannathan, and experimental assistance from N.~B. Bern.

\textbf{Author Contributions} M.W.R., S.K., and D.S.H.\ developed and conducted the experiments, after which M.W.R. and D.S.H. analysed the data. E.R. performed the numerical simulations under the guidance of M.M., who also developed the gauge transformations presented in Supplementary Material. Interactive feedback between the experiments and simulations carried out by M.W.R., D.S.H., E.R., and M.M.\ was essential to achieving the reported results. All authors discussed both experimental and theoretical results and commented on the manuscript.

\textbf{Author Information} Reprints and permissions information is available at \\ \url{www.nature.com/reprints}. The authors declare that they have no competing financial interests. Correspondence and requests for materials should be addressed to D.S.H.\ (dshall@amherst.edu).

\begin{center}
\textbf{METHODS}
\end{center}

\textbf{Condensate Production.} Condensates are produced in the $\ket{F{=}2,m{=}2}$ spin state of $^{87}$Rb by sequential steps of evaporative cooling, first in a time-averaged, orbiting potential (TOP) magnetic trap and subsequently in a $1,064$~nm crossed-beam optical dipole trap. The two evaporation stages are used to avoid the introduction of vortices that occasionally arise during the transfer of a condensate from the magnetic trap to the optical trap. The radial and axial optical trap frequencies at the end of the evaporative cooling process are $\sim 110$~Hz and $\sim 130$~Hz, respectively. A subsequent microwave Landau--Zener sweep drives the condensate into the $\ket{F{=}1,m{=}1}$ state.

After having established $B_z=10$~mG, the trap frequencies are increased to $(\omega_r,\omega_z)\approx 2\pi\times(160,220)$~Hz before the quadrupole field is turned on. The tighter trap better resists the magnetic forces exerted by the field gradient, but it also limits the condensate lifetime to approximately 500~ms as a result of three-body loss processes.  At the end of the experiment $N \approx 1.6 \times 10^5$, indicating typical three-body losses of $\sim 10\%$ .

\textbf{Magnetic Field Control.} The $x$, $y$, and $z$ axes are defined by the orientation of the magnetic field coils BX, BY and BZ, as shown in Fig.~\ref{fig:monopole-overview}{e}. The magnetic fields are calibrated to within $\sim 1$~mG using Majorana spectroscopy, in which the field component along the $z$~axis is rapidly (15~G s$^{-1}$) reversed and the fraction of atoms thereby transferred non-adiabatically to the $\ket{{-}1}$ state is measured as a function of the currents applied to the BX and BY field coils. Maximum transfer occurs when the transverse field components are minimised. The field along the $z$~axis is similarly calibrated by rapidly reversing the field component along the $x$~axis.

Precise magnetic field control at the location of the condensate is one of the most challenging aspects of the experiment. The condensate presents a small target ($\approx 7~\micron$) into which the field zero must be guided. The creation process is therefore quite sensitive to drifts in the relative position of the optical trap and the position of the field zero, limiting our ability to generate large sequential data sets without compensatory adjustments. Such drifts may be caused either by fluctuating background fields or by mechanical instabilities in the trapping beam optics. With $\Bqprime=3.7$~G~cm$^{-1}$, a 1-mG change in the radial field corresponds to a translation of the zero by 1.4~\micron, or 25\% of the condensate radius --- enough to disturb the creation of the monopole. Similar shifts alter the vertical bias field required to bring the field zero into the condensate.

An additional complication is that the centre of the optical trap and the physical centre of the gradient coils do not in general coincide, and can drift with respect to one another. The condensate can be offset horizontally by as much as 14~\micron, and vertically up to $\approx 25$~\micron below, the centre of the gradient coils.

\textbf{Adiabatic Spin Rotation.} As $B_z$ changes during the monopole creation process, the condensate spin ideally remains in the strong-field seeking state (SFSS), that is, the minimum-energy eigenstate of the local Zeeman Hamiltonian. At the field zero, however, the local Zeeman term of the Hamiltonian vanishes and non-adiabatic spin transitions to the neutral and weak-field seeking states become inevitable. Neglecting the kinetic-energy related to spin rotations and the weak spin--spin interactions in the condensate, the spatially-dependent probability of successful adiabatic spin rotation when the homogeneous bias field is inverted from large positive values to large negative values, can be approximated within the three-level Landau--Zener model by~\cite{Carroll1985}
\begin{align}
\Pad(x,y) =\left\{1-\exp\left[- \frac{\pi \mu_B \Bqprime^2 (x^2+y^2)}{4\hbar |\Bzdot|} \right]\right\}^2  \label{eq:LZ}
\end{align}
where $\mu_B$ is the Bohr magneton. The fraction of particles remaining in the SFSS can be approximated by an average of equation~\eqref{eq:LZ} weighted by a fixed particle density $\bar{n}({\bf r})$ as
\begin{align}
P_\textrm{ad}= \frac{\int\bar{n}({\bf r})\Pad({\bf r})\textrm{d}{\bf r}}{\int\bar{n}({\bf r})\textrm{d}{\bf r}}\label{eq:Pad}
\end{align}
Applying equations~\eqref{eq:LZ} and~\eqref{eq:Pad} to the initial vortex-free density distribution determined by solving the Gross--Pitaevskii equation with the parameter values extracted from the experiments, we obtain $P_\textrm{ad}=98\%$. The doubly-quantized vortex generated during the field inversion reduces the number of atoms in precisely the region where the undesired spin flips are most probable. For a density distribution that includes the doubly quantized vortex along the $z$~axis, equations~\eqref{eq:LZ} and~\eqref{eq:Pad} yield $P_\textrm{ad}=99\%$. Full numerical simulations of the creation of the doubly quantized vortex confirm that $99\%$ of the particles remain in the SFSS.

Experimentally, $P_\textrm{ad}$ is controlled by $\Bqprime$ and $\Bzdot$. Increasing $\Bqprime$ results in stronger magnetic forces on the condensate due to the gradient, which must remain small relative to those exerted by the optical trap so as not to perturb the condensate position extensively. The strength of the optical trap, however, cannot itself be increased without compromising both the size of the condensate and its lifetime. Choosing $\Bqprime=3.7$~G~cm$^{-1}$ was found convenient in this respect.

Decreasing $\Bzdot$, on the other hand, results in lengthier exposures of the BEC to magnetic field noise that can possibly induce undesirable spin transitions. The noise associated with the power mains (at frequencies that are odd integer $n$ multiples of 60~Hz) is the most serious, being resonant at a field of $n \times 85~\micro$G. The choice $\Bzdot=-0.25$~G/s ensures that the resonance condition is passed faster than a single oscillation period of the noise at least up to $n=7$. The effect of this noise is merely to distort the path traced by the field zero slightly during the creation ramp.

With the experimental parameters described above, we find that a negligible number of atoms is excited out of the SFSS after the field zero is moved fully through the condensate. Only when we increase the ramp rates by an order of magnitude do we find a discernible fraction of the atoms in the weak-field seeking state. This is consistent with the simulations and the calculation of $P_\mathrm{ad}$. We conclude that non-adiabatic spin flips are  not important in the monopole creation process with the parameters employed in the experiments, and that the Landau--Zener model describes this phenomenon well.

\textbf{Imaging.} At the end of the creation ramp, $B_z$ is rapidly decreased (in $0.040$~ms) until $|B_z/\Bzf| \gg 1$, a stage we call the `projection ramp'. This non-adiabatic field ramp keeps the order parameter essentially unchanged but takes the spin states $\{\ket{m}\}$ to be the approximate eigenstates of the Zeeman Hamiltonian. As described below, we image the particle density in each of these new eigenstates, accessing the detailed structure of the monopole established by the creation ramp.

Immediately after the projection ramp, the magnetic field gradient is turned off in $0.350$~ms. The optical trapping beams are then extinguished, releasing the condensate from the trap and permitting it to expand freely for 4~ms. The field is then increased adiabatically (in 1~ms) to 13.7~G in the $x$-direction as $B_z$ is simultaneously reduced to zero. After a 1.5~ms delay, the magnetic gradient coils are pulsed on for 3.5~ms to 20.1~G~cm$^{-1}$ (radial) to separate the spin states horizontally.

The total time of flight of the atoms is 22.9~ms, counted from the moment of release from the optical trap. After expansion, the condensates are imaged absorptively along both the vertical ($z$) and horizontal ($y$) axes simultaneously (to within $14~\micro{s}$) in the presence of a 0.1-G imaging field directed along $z$. In the absorption images we correct for neither the slightly different sensitivities of the different spin states to the probe beam nor the slightly different expansions that result from the applied magnetic field gradient.

Although we describe in this paper the creation of the monopole with the initial parameters $\Bqprime > 0$, $B_z(t=0) > 0$ and $\Bzdot <0$, the process yields essentially identical results experimentally when $\Bqprime > 0$, $B_z(t=0)<0$, and $\Bzdot > 0$ in the creation ramp, except that the field zero enters the condensate along the negative $z$~axis and thereby changes the sense of the spin rotation. Similarly, $B_z$ can be rapidly increased in either the $+z$ or $-z$ directions in the projection ramp with the same outcomes (Extended Data Fig.~\ref{fig:extendeddata}).

The images shown in Figs~\ref{fig:vortex_monopole_depth} and~\ref{fig:vortex_monopole-center} are selected from among dozens of similar images taken under identical conditions, and hundreds of images taken under similar conditions (Extended Data Fig.~\ref{fig:extendeddata}). Not every image demonstrates the signature presence of a monopole, because drifts in the magnetic fields and the location of the optical trap eventually cause the magnetic field zero to miss the condensate. Under optimal conditions we find that we can take five to ten sequential images in which the field zero passes through the condensate, after which we must adjust the magnetic bias fields to re-centre the magnetic field zero on the condensate.

The first images of pairs of singly-quantized vortices that indicate the passage of the monopole through the condensate were taken on February 6, 2013. Consistent images of the condensate density distributions associated with the monopole were first obtained on March 1, 2013.

\textbf{Numerical Simulation.} The experimental setup is simulated by solving the full three-dimensional Gross--Pitaevskii equation. The simulation parameters are chosen to match those of the experiment, but we  include neither the effects of three-body losses nor the magnetic forces arising from the gradient during the spin component separation just before imaging.  The particle number is held fixed at $N = 1.8\times10^5$ corresponding to the initial number of atoms in the experiment.  We can roughly account for the three-body losses by scaling the obtained particle density by the fraction of atoms that remain at the end of the experiment. Otherwise, the simulations are performed with the time-dependent parameters identical to those used in the experiment.

The volume considered varies from $20\times 20\times 20 \, a_r^3$ to $320 \times 320\times 320 \, a_r^3$, where $a_r=\sqrt{\hbar/(M\omega_r)}\approx 0.9~\micron$ is the radial harmonic oscillator length. The size of the computational grid changes from $180\times 180\times 180$ to $1,024\times 1,024\times 1,024$ points. The initial spin-polarised state is obtained with a relaxation method and the temporal evolution is computed using a split-operator technique employing Fourier transforms for the kinetic-energy part. The time required for the computation is reduced with the help of graphics processing units, coordinate transformations and an adaptive computational grid.

\textbf{Effects of free expansion.} The condensate must be allowed to expand freely in order to image its spin structure and determine the presence of the monopole. The condensate is therefore not imaged while the magnetic field zero is within the condensate. To demonstrate the effects of the free expansion on the spin structure, we show the simulated particle densities of the condensate just after the creation ramp in Extended Data Fig.~\ref{fig:supplement-simulation}. The images are created from an intermediate step in the complete simulation that is used to produce Fig.~\ref{fig:vortex_monopole_depth}. The principal effects of the release of the condensate are to cause it to expand with different speeds in different directions, to increase the relative vortex core sizes, to partially fill the vortex cores with other spin components, and the slight separation of the different spinor components. The last three effects are due exclusively to the repulsive interactions between the atoms during the first few milliseconds of expansion. Because there is excellent agreement between the simulated and experimentally observed results in Fig.~\ref{fig:vortex_monopole-center}, we conclude that Extended Data Fig.~\ref{fig:supplement-simulation} is a suitable representation of the condensate just after the creation ramp, while the field zero is still within the superfluid.


\clearpage
\renewcommand{\figurename}{Figure}

\clearpage

\clearpage

\clearpage

\clearpage

\setcounter{page}{1}
\renewcommand*{\thepage}{ED--\arabic{page}}

\stepcounter{myfigure}
\newcommand{\myfont}[1]{}
\renewcommand{\figurename}{Extended Data Figure}

\clearpage

\clearpage

\clearpage

\stepcounter{myfigure}
\renewcommand{\figurename}{Supplementary Information Figure}



\begin{center}
{\large\textbf{Observation of Dirac Monopoles in a Synthetic Magnetic Field}}\par
{\normalsize\textbf{Supplementary Information}}
\end{center}


\setcounter{page}{1}

\renewcommand*{\thepage}{SI--\arabic{page}}

\textbf{I. Spin Components} --- Our starting point is the spinor order parameter
\begin{align}\label{eq:orderp}
\mathit{\Psi}(\mb{r'},t) = \psi(\mb{r'},t)\zeta(\mb{r'},t) = \sqrt{n(\mb{r'},t)}e^{i\phi(\mb{r'},t)}\zeta(\mb{r'},t)
\end{align}
where $\psi$ is a scalar order parameter expressed in terms of the condensate particle density $n=|\psi|^2$ and phase $\phi=\arg(\psi)$, and $\zeta$ is a normalised spinor with $\zeta^\dagger\zeta=1$. The condensate spin is given by $\mb{S}=\zeta^\dagger\mb{F}\zeta$, where $\mb{F}=F_x\uv{x}'+F_y\uv{y}'+F_z\uv{z}'$ and $\{F_k\}$ are the usual spin-1 matrices.

As $B_z$ is decreased the spin rotates by an angle $\twiddle{\theta}=\pi - \theta'$ about an axis $\uv{n} = - \uv{x}'\sin\varphi' + \uv{y}'\cos\varphi'$ as illustrated in Fig.~\ref{fig:monopole-overview}. In the spinor basis $\ket{F,m}$ this spin rotation corresponds to a transformation matrix
\begin{align}
\mathcal{R}(\twiddle{\theta})\equiv e^{-i\mb{F}\cdot\uv{n}\,\twiddle{\theta}/\hbar} =
\begin{pmatrix}
\frac{1}{2}(1+\cos\twiddle{\theta}) & -\frac{1}{\sqrt{2}}e^{-i\varphi'}\sin\twiddle{\theta} & \frac{1}{2}e^{-2i\varphi'}(1-\cos\twiddle{\theta}) \\
\frac{1}{\sqrt{2}}e^{i\varphi'}\sin\twiddle{\theta} & \cos\twiddle{\theta} & -\frac{1}{\sqrt{2}}e^{-i\varphi'}\sin\twiddle{\theta} \\
\frac{1}{2}e^{2i\varphi'}(1-\cos\twiddle{\theta}) & \frac{1}{\sqrt{2}}e^{i\varphi'}\sin\twiddle{\theta} & \frac{1}{2}(1+\cos\twiddle{\theta})
\end{pmatrix}
\end{align}
The action of $\mathcal{R}$ on the initial spinor $(1,0,0)^\mathrm{T}$ is
\begin{align}\label{eq:rotatedspinor}
\mathcal{R}(\twiddle{\theta}) \begin{pmatrix}1 \\ 0 \\ 0 \end{pmatrix} =
\begin{pmatrix}
\frac{1}{2}(1+\cos\twiddle{\theta}) \\
\frac{1}{\sqrt{2}}e^{i\varphi'}\sin\twiddle{\theta} \\
\frac{1}{2}e^{2i\varphi'}(1-\cos\twiddle{\theta})
\end{pmatrix}
=
\begin{pmatrix}
\frac{1}{2}(1-\cos\theta') \\
\frac{1}{\sqrt{2}}e^{i\varphi'}\sin\theta' \\
\frac{1}{2}e^{2i\varphi'}(1+\cos\theta')
\end{pmatrix}=\zeta
\end{align}
From the $\varphi'$-dependence of the resulting spinor, we observe that the $\ket{{-}1}$ component contains a doubly-quantized vortex, and the $\ket{0}$ component contains a singly-quantized vortex, and all three components vanish along a nodal line that lies along the $+z'$~axis.

\textbf{II. Dirac Monopole in $\Bfield$} --- The gauge-field theory for spinor BECs has been published elsewhere~\citesi{Ho1996,Kawaguchi2012}; we reproduce here only its essential elements, following ref.~\citensi{Kawaguchi2012}. 
We begin with the condensate order parameter as expressed in equation~\eqref{eq:orderp}. The nonlinear Schr\"odinger equation governing the evolution of $\mathit{\Psi}$ may be recast as a Schr\"odinger equation for a charged particle described by a scalar field $\psi$ in the presence of synthetic electromagnetic scalar and vector potentials
\begin{align}\label{eq:Phizeta}
\mathit\Phi(\mb{r'},t)=-i\zeta^\dagger\frac{\partial\zeta}{\partial t}
\end{align}
and
\begin{align}\label{eq:afield}
\Afield(\mb{r'},t)=i\zeta^\dagger\del'\zeta
\end{align}
where $\del'$ indicates differentiation with respect to the primed coordinate system. These potentials lead to the synthetic electromagnetic fields
\begin{align}
\Efield = \hbar\left(-\del' \mathit\Phi - \frac{\partial \Afield}{\partial t}\right) \label{eq:Efield}
\end{align}
and
\begin{align}
\Bfield = \hbar\left(\del' \times \Afield \right) \label{eq:Bfield}
\end{align}
which, together with the Lorentz force law
\begin{align}
\frac{D}{Dt}(M \mb{v_s}) = \Efield + \mb{v_s} \times \Bfield - \del' \mulocal\label{eq:Lorentz}
\end{align}
describe the superfluid dynamics. In equation~\eqref{eq:Lorentz}, $D/Dt$ is the material derivative, $M$ is the atomic mass,
\begin{equation}
\mb{v_s}=\frac{\hbar}{M}\left(\del' \phi - \Afield\right) \label{eq:superfluidv}
\end{equation} is the superfluid velocity, and $\mulocal$ is a local chemical potential arising from kinetic, trap, and mean-field energies~\citesi{Kawaguchi2012}.

Applying equation~\eqref{eq:afield} to the specific spin rotation in our experiment (equation~\eqref{eq:rotatedspinor}), we find
\begin{align}\label{eq:Aspecific}
\Afield = - \frac{1 + \cos\theta'}{r' \sin \theta'} \mb{\hat{\boldsymbol{\varphi}'}}
\end{align}
If $\del' \phi = 0$, a typical initial condition for the trapped condensate, then $\mb{v_s} = -\hbar \Afield / M$, establishing equation~\eqref{eq:velocity}.

Equation~\eqref{eq:Aspecific} represents a vector potential with a singularity that lies along the nodal line in $\mathit{\Psi}$, that is, along the $+z'$~axis. Before calculating $\Bfield$ from this $\Afield$, consider the transformation
\begin{align}
\psi \rightarrow \twiddle{\psi} = \psi e^{-i\vartheta(\mb{r'},t)} \qquad \text{and} \qquad \zeta \rightarrow \twiddle{\zeta} = e^{i \vartheta(\mb{r'},t)} \zeta
\end{align}
where $\vartheta$ parameterises the transformation. The condensate order parameter $\mathit{\Psi} = \psi \zeta = \twiddle{\psi}\twiddle{\zeta}$, as well as all of its associated observables (including \Efield and \Bfield as implied by equation (\ref{eq:Lorentz})), are invariant under this transformation. Using equations~\eqref{eq:Phizeta} and~\eqref{eq:afield}, the synthetic scalar and vector potentials transform as
\begin{align}
\mathit{\Phi} \rightarrow \,\,\Phitilde \,\,= \mathit{\Phi} + \frac{\partial \vartheta}{\partial t}\label{eq:gaugePhi}
\end{align}
and
\begin{align}
\Afield \rightarrow \Atwiddle = \Afield - \del' \vartheta \label{eq:gaugeA}
\end{align}
respectively. These equations are recognised as those of a gauge transformation in ordinary electrodynamics. The invariance of the physical observable $\mb{v_s}$, and hence $\Vorticity$, with respect to choice of gauge may be confirmed by evaluating equation~\eqref{eq:superfluidv}.

One can use the choice of gauge to select the location of the singularity in $\Afield$. For example, the gauge function $\vartheta=-2\varphi'$ gives
\begin{align}
\Atwiddle = \frac{1-\cos\theta'}{r' \sin \theta'} \mb{\hat{\boldsymbol{\varphi}'}}
\end{align}
which has a singularity along the negative $z'$~axis. Similarly, choosing $\vartheta = -\varphi'$, yields
\begin{align}
\Atwiddle = -\frac{\cot \theta'}{r'}  \mb{\hat{\boldsymbol{\varphi}'}}
\end{align}
which contains two singularities that meet at the monopole, lying along the positive and negative $z'$~axes. Finally, by choosing $\vartheta = -2 \varphi'\mathit\Theta(z')$, where $\mathit\Theta(z')$ is the Heaviside step function, we obtain
\begin{align}\label{eq:Afield_nostring}
\Atwiddle = \frac{1}{r' \sin \theta'}\left[(1-\cos\theta')\mathit\Theta(z') - (1+\cos\theta')\mathit\Theta(-z')\right]\mb{\hat{\boldsymbol{\varphi}'}} + 2 \varphi' \delta(z') \uv{z'}
\end{align}
where $\delta(z')$ is the Dirac delta function. In this last case, the magnetic field is given directly by equation~\eqref{eq:monopolefield} without any singularities, i.e., the gauge transformation completely annihilates the Dirac string in $\Bfield$.

In the usual treatment of the Dirac monopole, the vector potential is expressed in terms of equation~\eqref{eq:Afield_nostring} without the last term~\cite{Wu1975}. Instead, two different gauges are employed for the first two terms, which can raise the question of how to connect, in practice, a solution extending across the gauge boundary. In contrast, we employ a single piecewise-defined gauge that answers this question and completely releases the Dirac monopole from its string in $\Bfield$. Thus a synthetic magnetic charge appears that is not strictly present in the gauge transformations involving the Dirac string in $\Bfield$.

We conclude that the singularity in $\Bfield$ is not physical, as it is inferred from the gauge-dependent location of the singularity in $\Atwiddle$. Note that this is not true for the vortex line singularity in $\Vorticity$, which depends on the location of the physical vortex line singularity in the gauge-invariant superfluid velocity $\mb{v_s}$ and must lie along the nodal line in $\mathit{\Psi}$.

\textbf{III. Faraday's Law} ---  In this section we use Faraday's law
\begin{align}
\del' \times \Efield =  - \frac{\partial \Bfield}{\partial t}\label{eq:Faraday}
\end{align}
to show explicitly that the motion of the monopole towards and through the condensate induces the superfluid velocity given by equation~\eqref{eq:velocity}. This approach is complementary to the preceding method of computing the superfluid velocity from the quasi-static spinor that aligns with the instantaneous applied magnetic field.

We assume that the monopole begins infinitely far away from the condensate on the positive $z'$~axis and moves towards it in the negative $z'$~direction. The synthetic electric field $\Efield$ resulting from the motion of the monopole may be written~\citesi{Kawaguchi2012}
\begin{align}
\Efield = \frac{\partial}{\partial t}(M \mb{v_s}) + \del' \left[\mulocal+\frac{1}{2}\,M \mb{v_s}^2\right]\label{eq:Kawaguchi}
\end{align}
The curl of this expression satisfies Faraday's law (equation~\eqref{eq:Faraday}), which in its integral form is
\begin{align}
\oint_\mathcal{C} \Efield \cdot d\mb{l} = -\frac{d}{dt}\int_\mathcal{S} \Bfield \cdot dS\label{eq:Faraday_int}
\end{align}
where the flux of $\Bfield$ is calculated through a capping surface $\mathcal{S}$ bounded by the directed curve $\mathcal{C}$.

We concentrate on the $\varphi'$ component of $\Efield$, which does not depend on the terms in the square brackets of equation~\eqref{eq:Kawaguchi} because those terms do not depend on $\varphi'$. On the other hand, the time-varying $\Bfield$ induces an $\Efield$ which in our case, for reasons of symmetry, is entirely in the $(\pm) \varphi'$ direction.

\begin{figure}[hb!]
\includegraphics[width=0.5\linewidth]{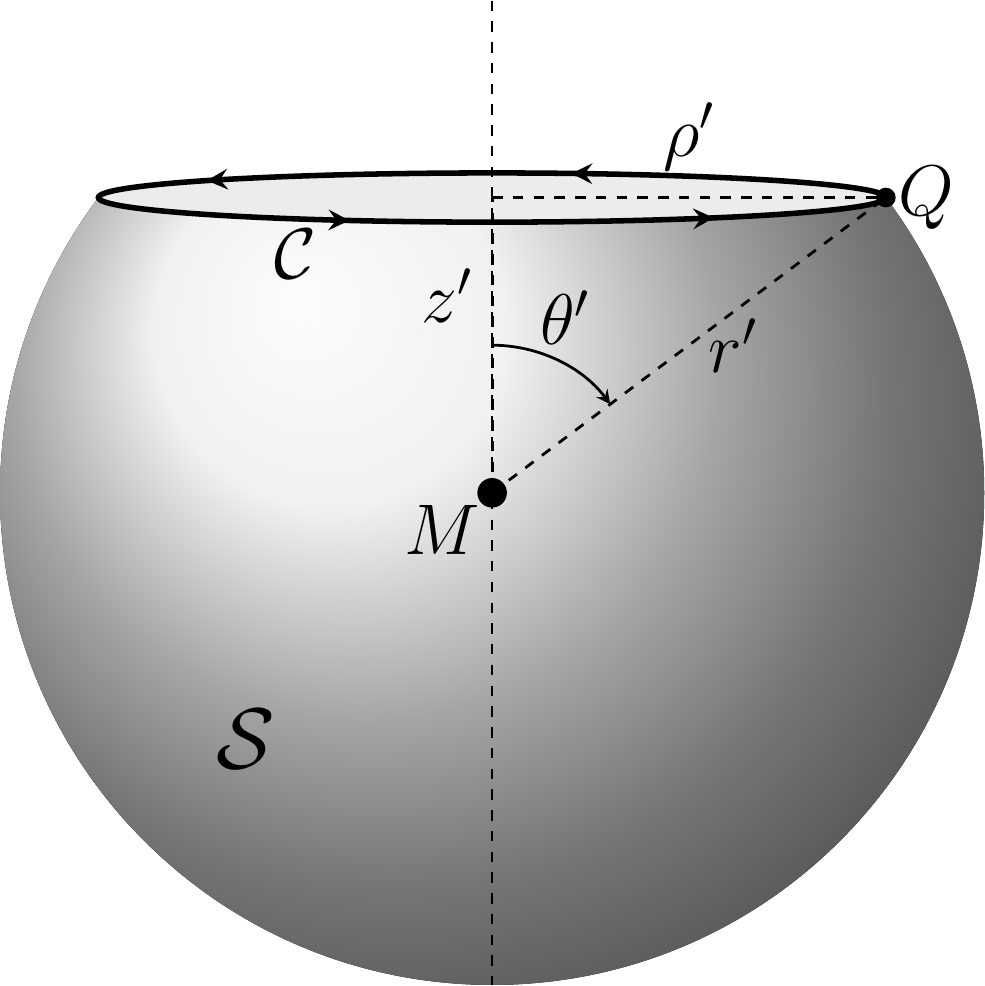} %
\caption{\label{fig:supplement-geometry}{\bf Calculation of the induced electric field due to the motion of the monopole.} The monopole is at $M$, and a representative superfluid mass element is at $Q$. The synthetic magnetic flux is calculated through the open surface $\mathcal{S}$, which is bounded by the curve $\mathcal{C}$. Faraday's law yields the $\varphi'$~component of \Efield at $Q$ as the monopole moves in the negative $z'$~direction (downwards).}
\end{figure}

Consider a superfluid mass element at $Q = (r',\theta',\varphi')$ with respect to the monopole (Supplementary Information Fig.~\ref{fig:supplement-geometry}). The curve $\mathcal{C}$ is a circular path of radius $\rho'=r'\sin\theta'$ passing through $Q$ and centred upon the $z'$~axis at a distance $z'=r'\cos\theta'$ from the monopole ($N$). Using equation~\eqref{eq:monopolefield} for the magnetic field of the monopole, equation~\eqref{eq:Faraday_int} yields
\begin{align}
\int_0^{2\pi} \Efield \cdot \rho'\,d\varphi'' = \frac{d}{dt} \left\{  \frac{\hbar}{r'^2} \int_{\theta'}^{\pi} r'^2 \sin \theta'' \,d\theta'' \int_0^{2\pi} d\varphi'' \right\}
\end{align}
where the flux integral is calculated over a fraction of the spherical surface $\mathcal{S}$ of radius $r'$ bounded by $\mathcal{C}$. Performing the integrals and simplifying, we find
\begin{align}
\Efield \cdot {\hat{\varphi}'} = \frac{d}{dt} \left\{ \frac{\hbar}{r'}\left( \frac{z' + r'}{\rho'}\right)\right\} = \frac{d}{dt} \left\{ \frac{\hbar}{r'}\left( \frac{\cos\theta' + 1}{\sin \theta'} \right)\right\}\label{eq:last}
\end{align} 
Equation~\eqref{eq:velocity} then follows directly by equating equation~\eqref{eq:last} with the $\varphi'$ component of equation~\eqref{eq:Kawaguchi}.

\bibliographystylesi{naturemag}
\bibliographysi{monopole}


\end{document}